\documentclass[conference]{IEEEtran}
\IEEEoverridecommandlockouts
\usepackage{cite}
\usepackage{amsmath,amssymb,amsfonts}
\usepackage{algorithmic}
\usepackage{graphicx}
\usepackage{textcomp}
\usepackage{xcolor}
\usepackage{caption}
\usepackage{subcaption}
\usepackage[hidelinks]{hyperref}
\usepackage{listings}
\usepackage{fancyvrb}
\def\BibTeX{{\rm B\kern-.05em{\sc i\kern-.025em b}\kern-.08em
    T\kern-.1667em\lower.7ex\hbox{E}\kern-.125emX}}
\begin{document}

\lstset{upquote=true}

\title{(Genetically) Improving Novelty in Procedural Story Generation
\thanks{This work has been supported by Grand Valley State University.}
}

\author{\IEEEauthorblockN{Erik M. Fredericks}
\IEEEauthorblockA{\textit{School of Computing} \\
\textit{Grand Valley State University}\\
Allendale, USA \\
frederer@gvsu.edu}
\and
\IEEEauthorblockN{Byron DeVries}
\IEEEauthorblockA{\textit{School of Computing} \\
\textit{Grand Valley State University}\\
Allendale, USA \\
devrieby@gvsu.edu}
}

\maketitle

\begin{abstract}
Procedural story generation (PCG) tailors a unique narrative experience for a player and can be accomplished via multiple techniques, from matching storylets to grammar-based generation.  There exists a rich opportunity for evolutionary algorithms to be applied to this domain for intelligently constructing game narratives.  This paper describes a conceptual procedure for applying genetic improvement to a grammar-driven procedural narrative within the context of a browser-based game.
\end{abstract}


\section{Introduction}

Procedural content generation typically focuses on developing unique experiences commonly found in genres such as roguelikes (e.g., Nethack, Dungeon Crawl Stone Soup, etc.\footnote{See \url{https://www.nethack.org/} and \url{http://crawl.develz.org/}, respectively.}), where intelligent algorithms leverage a set of rules and/or random chance to instantiate varying situations.  Such algorithms can lead to \textit{emergent behaviors} that can enable  a rich player experience, where such behaviors are not necessarily hard-coded by the developers. 

Story generation tends to be a difficult aspect of procedural content generation, as simply combining snippets or paragraphs does not generally lead to a cohesive narrative.  Narratives can be automatically crafted using trees or grammars with storylets (i.e., small pieces of narrative) that are ``known'' to work well together.  Consider the following storylet: ``You enter a new room.  It is filled with lush vegetation and is quite humid.  A large crustacean lounges in the corner.''  Each sentence within the storylet can be procedurally generated using a large bank of phrases with meta-data to enable precise matching.  For instance, the storylet can be represented as a grammar, where words surrounded by \# represent future production rules: 

\begin{lstlisting}[basicstyle=\small]
S $\mapsto$ #newRoom#.  It is #verb#.ed with
  #adjective# #noun#. #randomOccurrence#. 
\end{lstlisting}

Additional meta-data checks would be required to ensure that a \textit{flow} exists as well to avoid confusing the player (e.g., it would not make sense for lush vegetation to exist within a snowy environment). This paper presents early efforts towards applying genetic improvement to procedural story narrative.  Specifically, we focus on \textit{diversifying} the set of possible cohesive narrative states generated via grammar-based narrative storytelling.  We next describe methods for procedural storytelling, our work in progress as an illustrative example, how we will apply genetic improvement to our story grammar, and summarize with a discussion.

\section{Procedural Storytelling}

There exist many frameworks for enabling narrative-driven games, including Twine and Ink.\footnote{See \url{http://twinery.org/} and \url{https://www.inklestudios.com/ink/}, respectively.}  Each are open source and provide an easy-to-use interface for creating text-driven games.  Games can be published as webpages or incorporated as modules in more advanced engines (e.g., Unity, Godot, Unreal Engine, etc.).  Mason \textit{et al}. recently introduced Lume, a tool for enabling procedural narrative generation via parameterized node trees that use bindings to maximize narrative coherency~\cite{mason.2019}.   Tracery is an open-source tool for creating generative text using a grammar-based system that has been used in multiple applications, including procedural narratives, Twitter bots, and role-playing game mechanics~\cite{compton.2015}.  Tracery has been ported to most common languages as well. 


\section{Work in Progress}

\begin{figure}[htbp]
\centering
\includegraphics[width=3.3in]{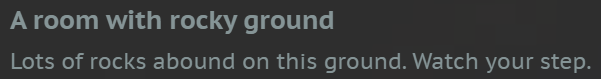}
\caption{Room title and description.}
\label{fig:title}
\end{figure}

\begin{figure}[htbp]
\centering
\includegraphics[width=3.3in]{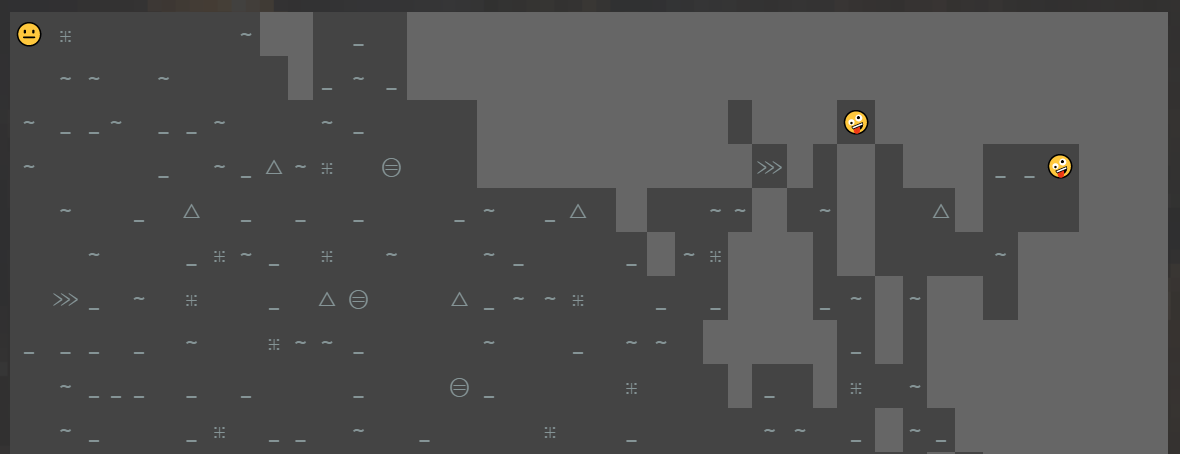}
\caption{Cropped minimap of environment.}
\label{fig:minimap}
\end{figure}

Figures~\ref{fig:title} and \ref{fig:minimap} present a sample of our procedurally-generated content.  Figure~\ref{fig:title} demonstrates a room title and description that has been generated in accordance with Simplex noise~\cite{perlin.2001} and a Tracery grammar~\cite{compton.2015}, and Figure~\ref{fig:minimap} illustrates the space around the player, where the emoji faces represent a player and non-player characters and other characters represent Simplex noise-generated environmental features (e.g., $\Delta$ is a tight tunnel and {\raise.17ex\hbox{$\scriptstyle\sim$}} is a stream).

Simplex noise is a technique commonly used for generating ``smooth'' environmental features in game maps~\cite{redblob.2020}.  For this work we are populating a two-dimensional grid with Simplex noise values, where each value translates to a specific environment feature.  For instance, a noise value within $[0.35, 0.55]$ results in a room with the meta-attribute \texttt{STREAM} to denote that a stream flows within the room, and we execute Tracery to generate a relevant storylet for a \texttt{STREAM}-tagged grammar.  We normalize our Simplex values between $[0.0, 1.0]$.

A sample Tracery grammar and sentence generation for a \texttt{STREAM} environment is as follows, where the grammar has been significantly reduced in scope for presentation purposes:

\begin{lstlisting}[basicstyle=\small]
var rules = {
  'origin':['[myPlace:#path#]#line#'],
  'path':['path', 'rock', 'cavern wall', 
  'line':['#stream.a.capitalize#'],
  'nearby':['beyond the #path#', 
    'far away', ...],
  'substance':['light', 'reflections', 
    'mist', 'shadow', 'darkness'], 
    'underfoot', 'stalagmites', ...],
  'stream': ['#stream-type# 
    #stream-verb.s# #nearby#'],
  ...
};
var g = tracery.createGrammar(rules);
console.log(g.flatten('#origin#'));
\end{lstlisting}

In this example, \texttt{rules} represents the Tracery grammar, \texttt{g} is the Tracery object, and \texttt{\#origin\#} represents the starting point of the grammar for Tracery to parse and generate a sentence.  Based on a set of grammars as illustrated, a large number of storylets can be generated via Tracery and its grammar constraints.  We next describe how novelty search will be applied as a genetic improvement technique for this case study.

\subsection{Genetic Improvement}

A major focus with this game environment is to provide as many diverse, yet cohesive, storytelling opportunities to the player as possible.  Therefore, we augment our set of Tracery grammars with grammatical evolution~\cite{oneill.2001}, where our fitness is replaced by novelty search~\cite{lehman.2004}.  Grammatical evolution focuses on evolving an individual (typically a program) via a grammar as opposed to a tree.  Novelty search is an evolutionary computation-based technique for finding as many diverse, yet still optimal, solutions to a given problem.  In general, novelty search uses similar evolutionary operations to genetic algorithms/grammatical evolution, however the fitness function is typically superceded by a novelty metric.   

%
%
%
%

The \textit{novelty metric} (i.e., a mathematical formula for encouraging diversity) is used within novelty search to guide the search process to distinct areas of the solution space.  For this application, we construct our novelty metric to fulfill the constraints of our Simplex noise-generated map while encouraging diversity via natural language processing (NLP) metrics.  While many advanced NLP algorithms exist for calculating differences between sentences (where a study of such algorithms can form the basis of future work)~\cite{al.2019}, we will use \textit{Word2Vec} for measuring similarity between sentences~\cite{mikolov.2013}.  We anticipate training \textit{Word2Vec} on open-source corpora.  Equation~\ref{eqn:pos} demonstrates usage of \textit{Word2Vec} for calculating sentence similarity (following training).

\begin{equation}
    sim(\mu_i, \mu_j) = word2vec.model(\mu_i, \mu_j),
    \label{eqn:pos}
\end{equation}

where $\mu_i$ and $\mu_j$ represent sentences generated from Tracery grammars. Equation~\ref{eqn:novelty} next shows our novelty score calculation:

\begin{equation}
novelty(k) = \frac{1}{k} \sum_{i=0,j=0,i \ne j}^{k} sim(\mu_i, \mu_j)
    \label{eqn:novelty}
\end{equation}

The most diverse solutions (i.e., passing a novelty threshold score) are added to a \textit{novelty archive} that is maintained throughout the course of the search procedure, where this archive is returned as output following search completion.  Additional tags may be necessary to ensure that generated solutions are \textit{feasible} from a grammatical perspective.

\section{Discussion}

This paper has introduced an early proof-of-concept for automatically improving novelty in procedurally-generated storylines for games.  The proof-of-concept uses Twine to prototype the game environment, Tracery for enabling grammar-based storylet generation, and Simplex noise for generating a cohesive environment.  We plan to further augment this prototype with a novelty-search based grammatical evolution heuristic for generating diverse Tracery grammars and Word2Vec to measure similarity.    

\bibliographystyle{IEEEtran}
\bibliography{IEEEabrv,efredericks_master}

\end{document}